\documentclass[letterpaper]{jpconf}
\usepackage{graphicx}
\begin{document}
\title{Search for Single Top Quark Production at the D0 Experiment using Bayesian Neural Networks}

\author{Andres J. Tanasijczuk, for the D0 Collaboration}

\address{University of Buenos Aires}

\ead{andres@fnal.gov}

\begin{abstract}
We present the methodology used to measure the single top quark production cross section in the D0 experiment, and show as an example the results that led to the first evidence of single top quark production in D0 at the Fermilab Tevatron proton-antiproton collider. The selected events are mostly backgrounds, which we separate from the expected signals using three multivariate analysis techniques, one of them being Bayesian neural networks, which we will describe here.
\end{abstract}

\section{Introduction}

At hadron colliders, top quarks are mostly produced in pairs via the strong interaction from the decay of a virtual gluon. Less frequently, they can also be produced singly via the electroweak interaction from the decay of a virtual $W$ boson. Although, the single top quark production cross section is only half of the top pair cross section, single top events are much more difficult to isolate from the backgrounds, to the point that a simple counting experiment can not be performed as it is usually done in top pair analyses. Therefore, more powerfull methods are required to separate the signal from the background. In D0, three multivariate techniques are used: Bayesian Neural Networks (BNN), Boosted Decision Trees (BDT) and Matrix Elementes (ME). We will focus on the BNN method. The analysis results presented here were published in 2008~\cite{PRD}, and are based on 0.9 fb$^{-1}$ of data collected by the D0 detector at the Tevatron Collider.

\section{Single Top Production, Event Selection and Backgrounds}

In the Tevatron there are two electroweak processes that contribute to single top quark production. These are the $s$-channel and the $t$-channel production modes.
In the $s$-channel the top quark is produced in association with a bottom quark, while in the $t$-channel a light quark is also produced. The respective Standard Model (SM) next-to-leading (NLO) order cross sections at 1.96 TeV are 1.04 $\pm$ 0.04 pb and 2.26 $\pm$ 0.12 pb for a top quark mass of $m_{t}$ = 175 GeV/c$^{2}$. We refer to the $s$-channel process as ``$tb$'' production, where $tb$ includes both $t\bar{b}$ and $\bar{t}b$ states, and to the $t$-channel process as ``$tqb$'' which includes the states $tq\bar{b}$, $t\bar{q}\bar{b}$, $\bar{t}\bar{q}\bar{b}$ and $\bar{t}qb$.

The final state event selection includes an isolated lepton (electron or muon) with high transverse momentum, missing transverse energy indicating the presence of a neutrino, and two, three or four jets, where one or two of them must to be tagged as $b$-jet, making a total of 12 independent analysis channels. The main backgrounds are $W$ bosons produced in association with jets, top-pair production with decay into lepton+jets and dilepton final states, and multijets production where one of the jets is misreconstructed as an electron or a heavy-flavor quark decays to a muon that passes the isolation criteria.

\begin{figure}[!h!tbp]
\includegraphics[width=0.32\textwidth]
{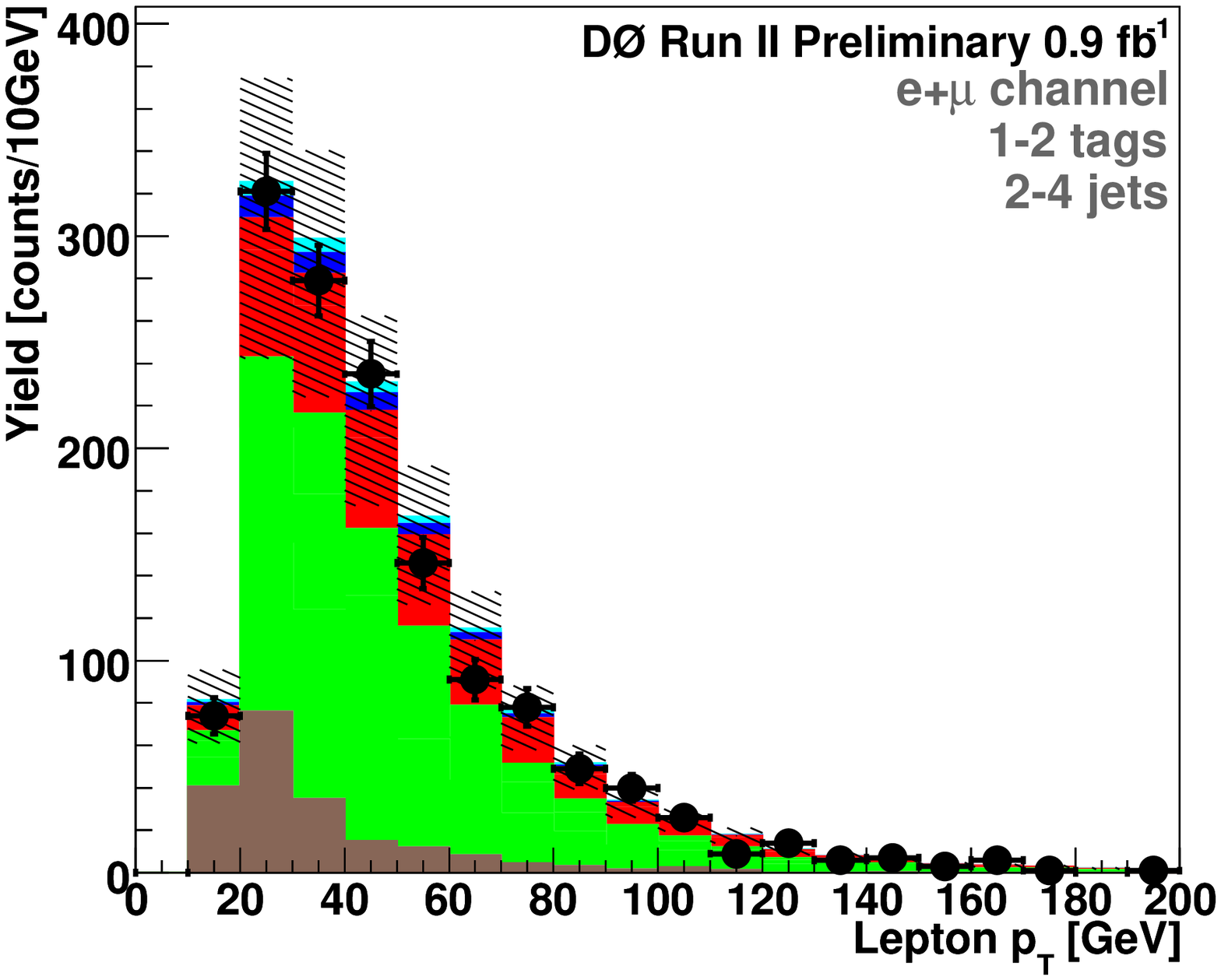}
\includegraphics[width=0.32\textwidth]
{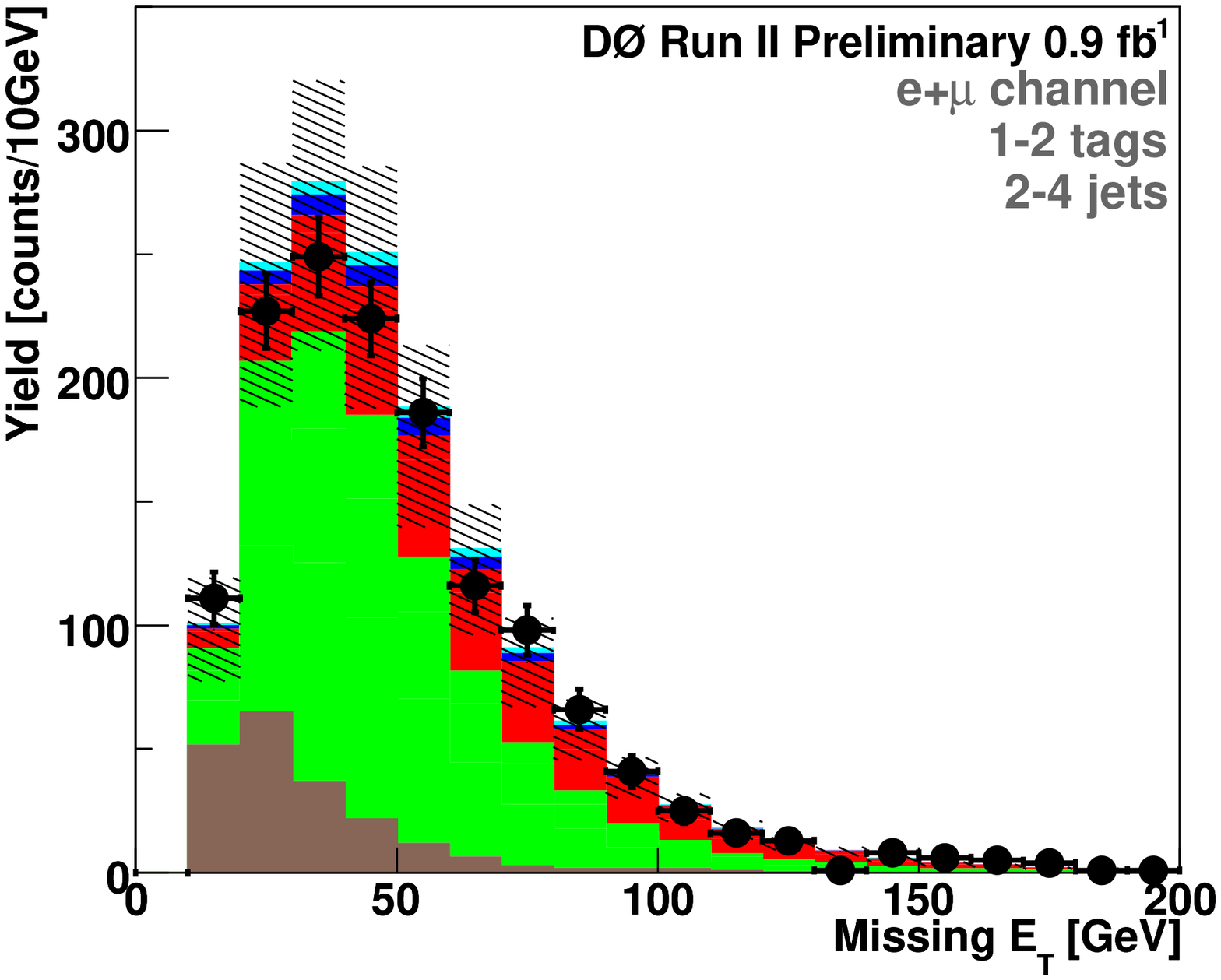}
\includegraphics[width=0.32\textwidth]
{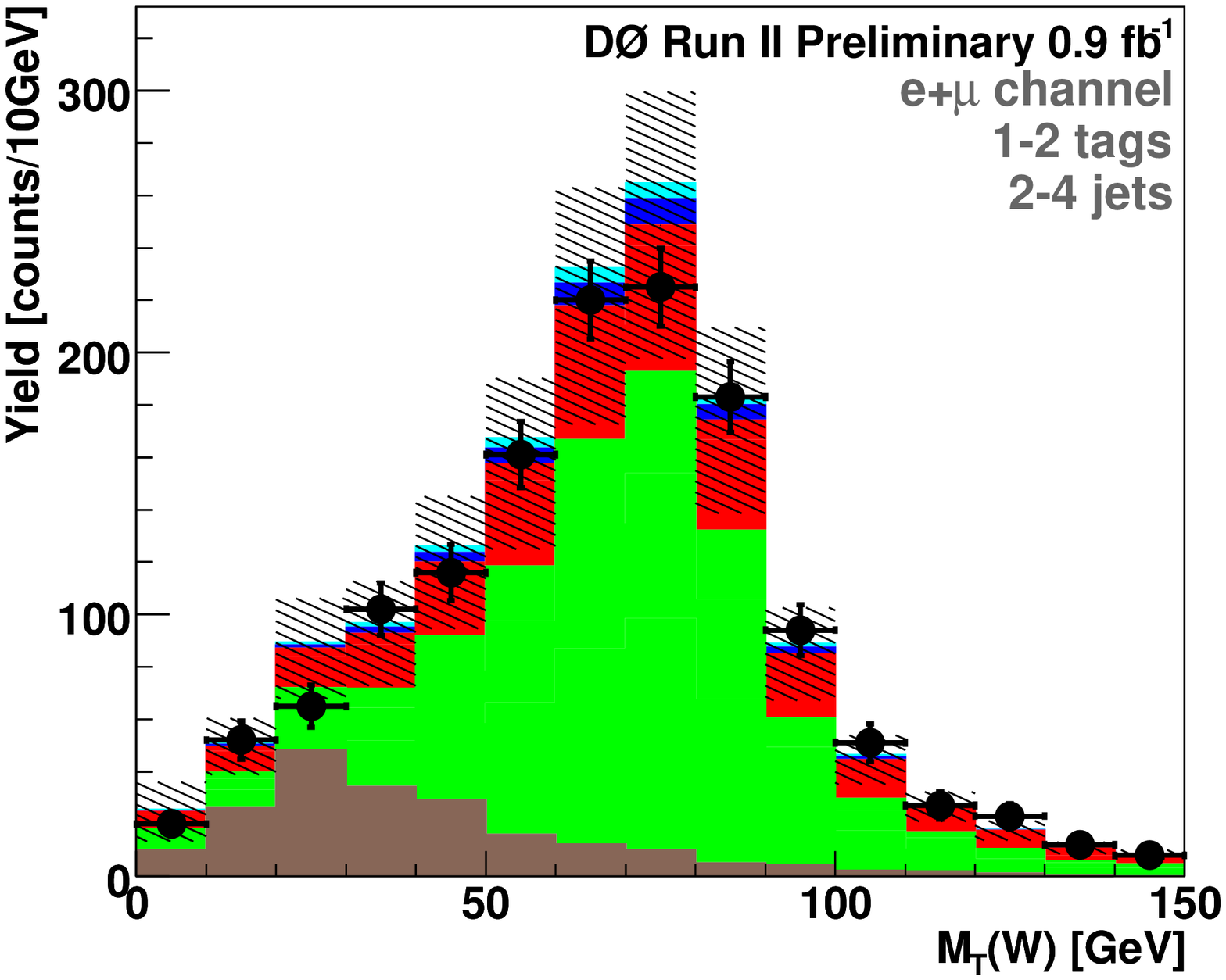}
\caption[pretag]{Data-Background comparison after event selection for three of the most important variables: the lepton transverse momentum (left), the missing transverse energy (centre) and the transverse mass of the $W$ boson (right). Backgrounds are grouped together in the three main components: fake-leptons (brown), $W$+jets (green) and top-pairs (red). Signal $tqb$ channel is shown in dark blue and signal $tb$ channel in light blue. The hatched area represents the one sigma uncertainty on the background.}
\label{fig:pretag}
\end{figure}

Figure~\ref{fig:pretag} shows a comparison between the data and the model of the sum of backgrounds plus signal after event selection. The hatched area in the histograms represents the one sigma uncertainty on the background. One can see that the expected signal is much smaller than this uncertainty, making a counting experiment impossible to be performed. Therefore the need to use more powerful techniques that exploit the kinematic differences between the signal and the background events.

\section{Multivariate Methods}

In D0, three multivariate techniques are used to separate further the signal from the background. Two of them are machine learning techniques (BNN and BDT), while the third one consists on matrix element calculations. What the three methods have in common is that they all use some set of input variables $\{x\}$ to approximate the discriminant $D(x)$,
which is the probability for an event to be signal. We will shortly describe only the BNN method.

\subsection{Bayesian Neural Networks}

From the structural point of view, a neural network is an interconected set of nodes. The network we use is divided in three layers. The first layer consists of $N_{var}$ input nodes, where $N_{var}$ is the number of input variables (typically between 20 and 25), the second layer consists of 20 hidden nodes, and the third layer has only one node, which will provide the output of the network (the discriminant). All the nodes in one layer are connected to all the nodes in the next layer.

As any machine learning technique, a neural network must to be trained. During the training process, the strenght (or in mathematical terms, the weight) of each node-to-node connection is determined in order to get the best possible signal from background separation. In order to avoid any bias, the training should be performed using samples that are independent of the samples that will be later used to build the discriminant. So we divide our signal and background MC samples in two subsets and use the first subset for the BNN training and the second subset to later compare the BNN output in data to the BNN output in the signal plus background model.

A Bayesian neural network is build taking a weighted average over many (100) neural networks trained iteratively. This allows one to average out statistical fluctuations.

To select the set of input variables for the BNN we first choose a set of well modeled variables (in this analysis approximately 60) using the $Kolmogorov$-$Smirnov$ test. Then we choose in each of the 12 analysis channels the subset of variables that have the highest signal-to-background separation power. To that end we use $Rulefit$~\cite{Rulefit}, a multivariate method based on the predictive learning via Rule Ensembles technique that gives as a side product a ranking of the input variables. The parameter that defines the ranking is called $importance$ and it ranges from 0 to 100. We keep those variables with importance higher than 10\% of the maximum ($importance >$ 10). The samples used for $Rulefit$ are the same as the ones used for the BNN training.

Once the BNN is trained, we look at the BNN output in the seond half of signal and background samples, expecting to see the signal events populating mostly the region near one and the background events being pushed towards zeroi. Toghether with other figures of merit we judge the BNN convergence and its performance. Then we apply the BNN discriminant to the data.


\section{Cross Section Measurement}

\begin{figure}[!h!tbp]
\includegraphics[width=0.32\textwidth]
{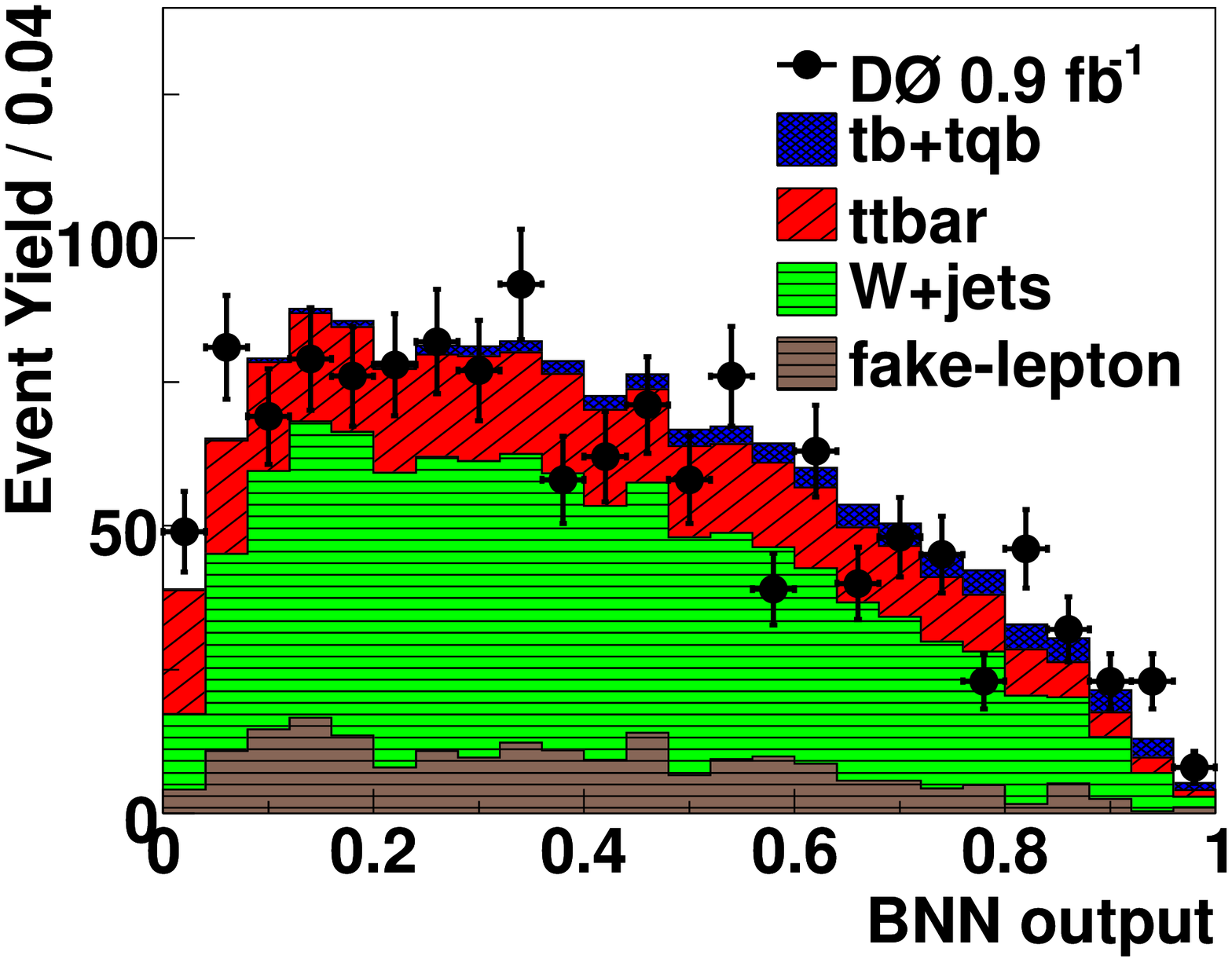}
\includegraphics[width=0.32\textwidth]
{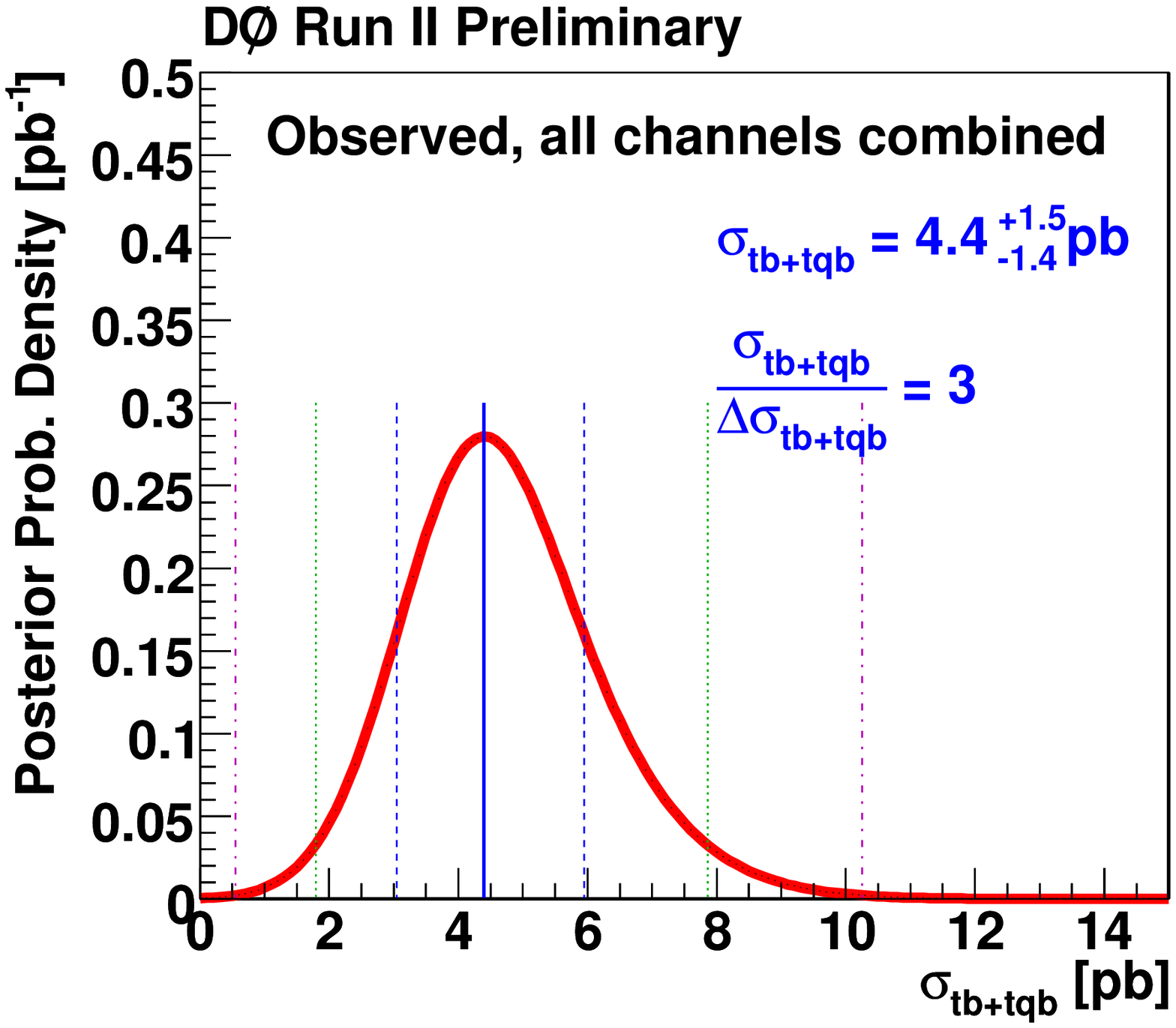}
\caption[BNNoutput_posterior]{Left: BNN output distribution fo rhte 12 analysis channels combined. Right: Posterior probability density for the single top quark production cross section for all the 12 analysis channels combined, derived from the BNN discriminant shown on the left plot.}
\label{fig:BNNoutput_posterior}
\end{figure}

The left plot in Figure~\ref{fig:BNNoutput_posterior} shows the BNN discriminant distribution for the 12 analysis channels combined. Good agreement is seen between the data and the background (plus signal) model. Signal appears mostly in the high discriminant region.

From this discriminant, a Bayesian posterior probability density distribution is computed as a function of the single top quark production cross section ($\sigma_{tb+tqb}$) using a binned likelihood approach including all of the systematics. For the previously mentioned BNN output, the posterior is shown in the right plot in Figure~\ref{fig:BNNoutput_posterior}. In the calculation, the $\sigma_{tb+tqb}$ prior is set to a constant non-negative function, representing our total ignorance of the value of $\sigma_{tb+tqb}$. We define the measured value of $\sigma_{tb+tqb}$ as the position of the peak of the posterior, and the uncertainty $\Delta\sigma_{tb+tqb}$ as the lower and upper limits of the 68.3\% two-sided confidence interval. The ratio $\sigma_{tb+tqb} / \Delta\sigma_{tb+tqb}$ gives an approximate estimate of the significance of the measurement.

\begin{figure}[!h!tbp]
\includegraphics[width=0.24\textwidth]
{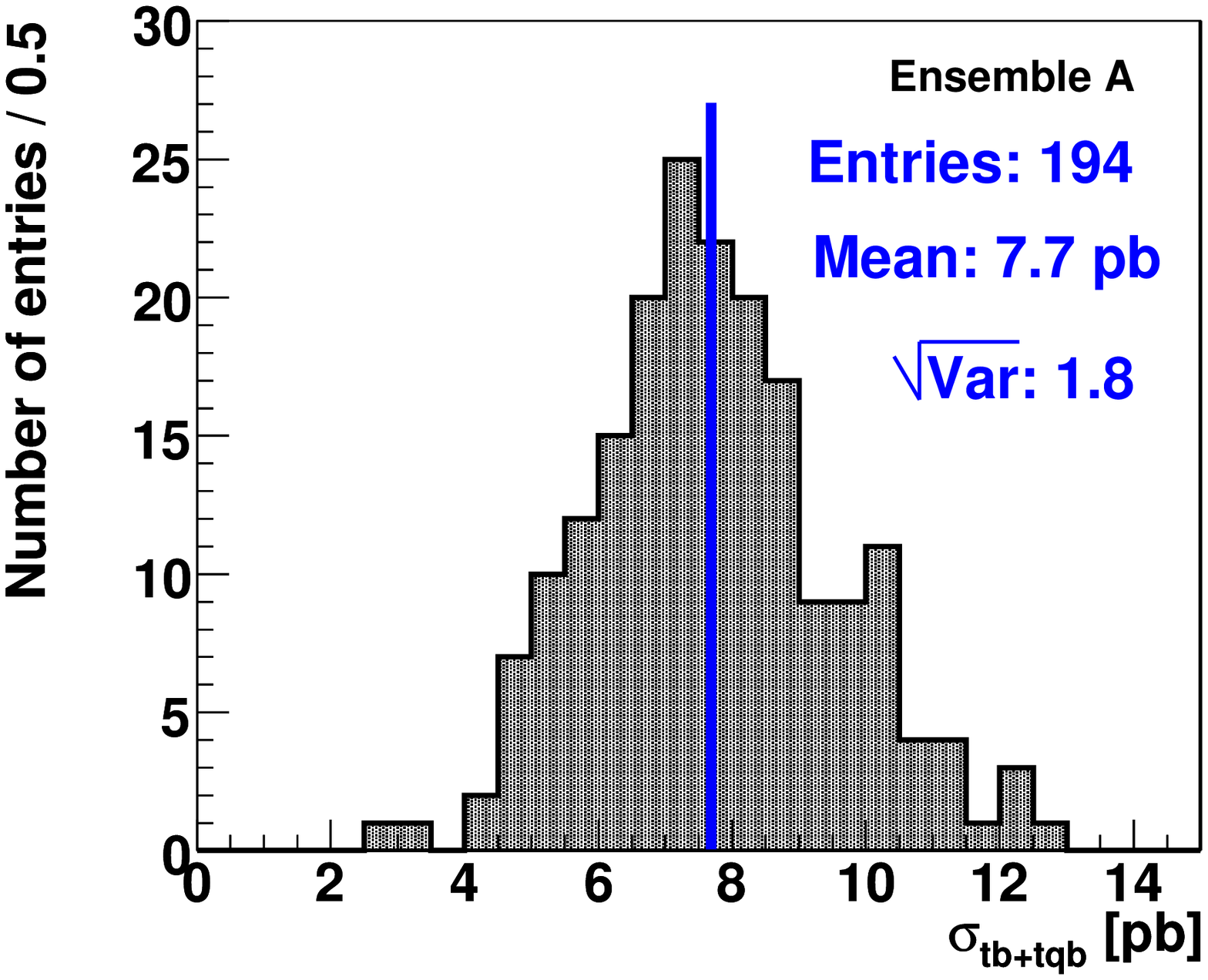}
\includegraphics[width=0.24\textwidth]
{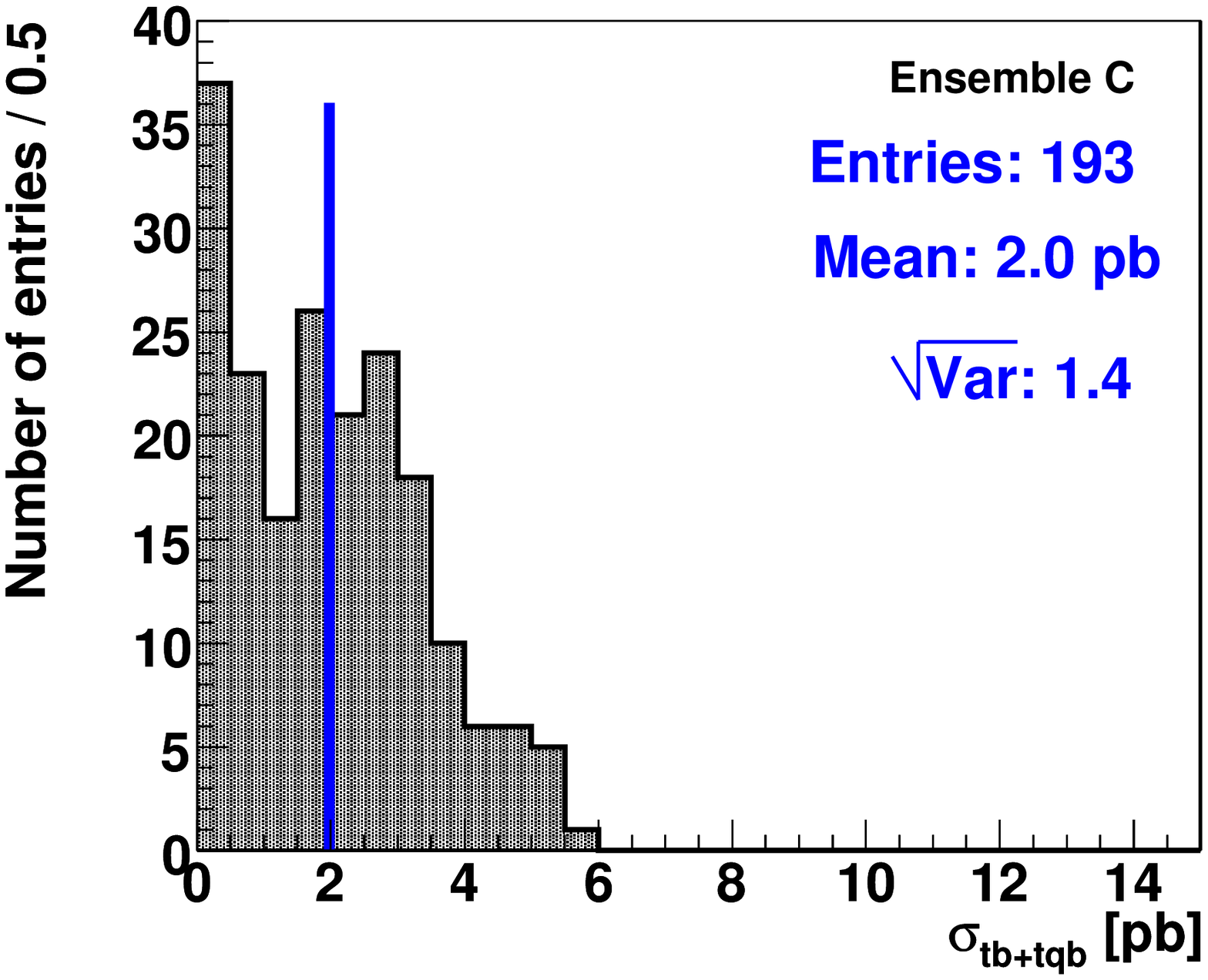}
\includegraphics[width=0.24\textwidth]
{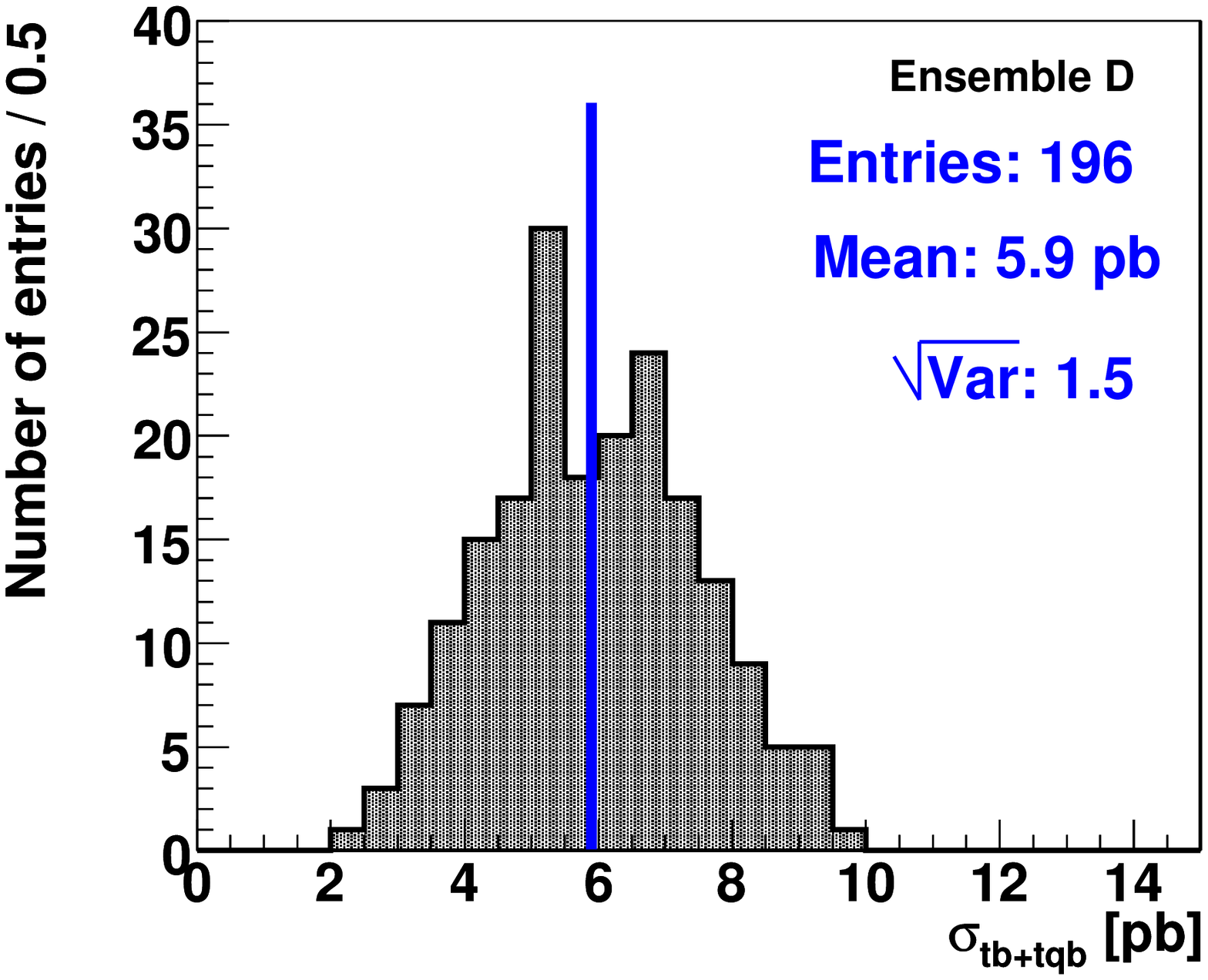}
\includegraphics[width=0.26\textwidth]
{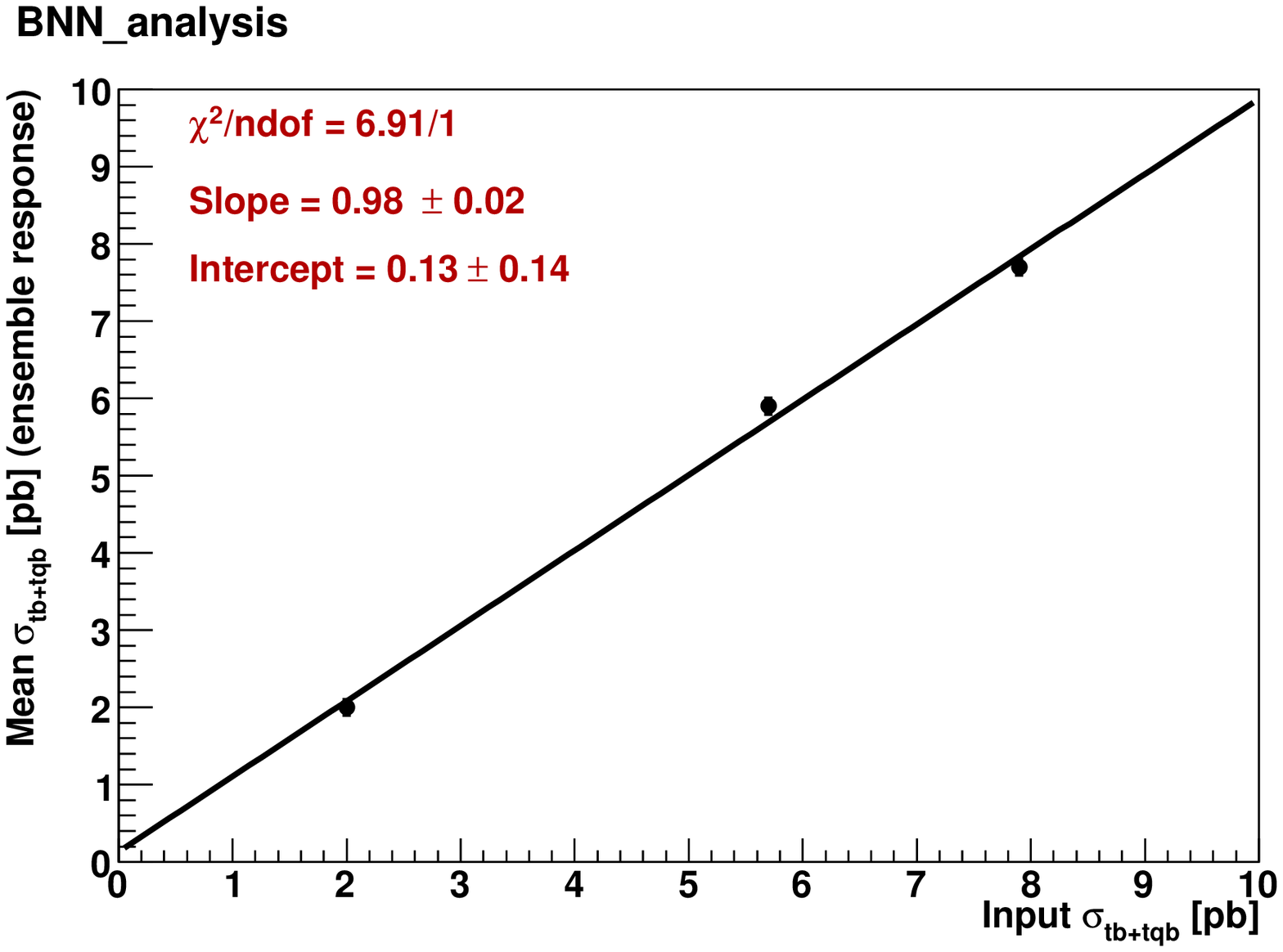}
\caption[BNN_calibration]{The three first plots on the left show the distributions of the measured $\sigma_{tb+tqb}$ in three ensembles of pseudo-datasets, each ensemble containing a different amount of $tb+tb$ signal (unknown to the analyzer). The right most plot shows the measured $\sigma_{tb+tqb}$ in the three ensembles of pseudo-datasets (shown in the three left plots) plotted against the input $\sigma_{tb+tqb}$. A linear fit to the points is also shown.}
\label{fig:BNN_calibration}
\end{figure}

In order to calibrate our cross section extraction method, we generate a few ensembles having each of them thousands of pseudo-datasets. Each pseudo-dataset is like one D0 experiment with 0.9 fb$^{-1}$ of ``data'' over which we ran the whole analysis. All pseudo-datasets in a given ensemble contain background events plus the same unknown (to the analyzer) amount of $tb+tqb$ signal in the SM ratio. We then make distributions of the measured cross section for each ensemble separately (first three left plots in Figure~\ref{fig:BNN_calibration}) and plot the mean values against the input $tb+tqb$ cross section (right plot in Figure~\ref{fig:BNN_calibration}). Then, a linear fit to the points is done. In Figure~\ref{fig:BNN_calibration} the slope of the fit is consistent with unity and the intercept consistent with zero, so no scale or shift were needed to apply to the cross section extracted from the posterior.

The measured inclusive single top quark production cross sections $\sigma(p\bar{p} \rightarrow tb + X, tqb + X)$ are $4.9+1.4-1.4$ pb for BDT, $4.4+1.6-1.4$ pb for BNN and $4.8+1.6-1.4$ pb for ME. Since the three analysis methods are not completely correlated, we combine them using another BNN. This BNN uses as input the discriminants of the three multivariate methods and gives its own discriminant output. For the combination, the measured single top quark production cross section is $4.7+1.3-1.3$ pb.

\section{Significance}

The final step in the analysis is to give the sensitivity of the measurement, which is equivalent to say by how much one is able to rule out the background-only hypothesis. The sensitivity is expressed giving the $p$-value, which represents the probability that the background could fluctuate up to give the measured cross section value or higher. To calculate the $p$-value we use an ensemble of pseudo-datasets that contain background but no signal. The $p$-value is given by the fraction of pseudo-datasets for which the measured cross section is at least as high as our result. Morover, assuming a gaussian distribution, one can translate the $p$-value to standard deviations (SD). If the $p$-value is calculated based on the SM single top cross section, one gets the expected sensitivity. They are 2.2, 2.1 and 1.9 SD for the BNN, BDT and ME analyses respectively. If on the other hand the calculation is based on the observed (measured) single top cross section, one gets the observed significance, which are 3.4, 3.2 and 3.1 SD for the BDT, ME and BNN analyses respectively. A significance greater than three SD is considered as evidence, so we can say that the three methods reached the evidence of single top quark production. The significances of the combined analysis are 3.2 SD expected and 3.6 SD observed.

\end{document}